\documentclass[aps,prl,twocolumn,showpacs,superscriptaddress,floatfix]{revtex4-1}
\usepackage[utf8]{inputenc}
\usepackage[gen]{eurosym}
\usepackage{amsmath,amssymb}
\usepackage{graphicx}
\usepackage{epsfig}
\usepackage{color}
\usepackage{rotating}
\usepackage{bm}
\usepackage{setspace}
\usepackage{hyperref}
\usepackage[none]{hyphenat}

\begin{document}
\title{Full-Gap Superconductivity in BaAs/Ferropnictide Heterostructures}
\author{Ming-Qiang Ren}
\affiliation{State Key Laboratory of Low-Dimensional Quantum Physics, Department of Physics, Tsinghua University, Beijing 100084, China}
\affiliation{Department of Physics, Southern University of Science and Technology, Shenzhen 518055, China}
\author{Qiang-Jun Cheng}
\affiliation{State Key Laboratory of Low-Dimensional Quantum Physics, Department of Physics, Tsinghua University, Beijing 100084, China}
\author{Hui-Hui He}
\affiliation{Department of Physics and Beijing Key Laboratory of Opto-electronic Functional Materials $\&$ Micro-nano Devices, Renmin University of China, Beijing 100872, China}
\affiliation{Key Laboratory of Quantum State Construction and Manipulation (Ministry of Education), Renmin University of China, Beijing 100872, China}
\author{Ze-Xian Deng}
\author{Fang-Jun Cheng}
\author{Yong-Wei Wang}
\author{Cong-Cong Lou}
\affiliation{State Key Laboratory of Low-Dimensional Quantum Physics, Department of Physics, Tsinghua University, Beijing 100084, China}
\author{Qinghua Zhang}
\affiliation{Institute of Physics, Chinese Academy of Sciences, 100190 Beijing, China}
\author{Lin Gu}
\affiliation{State Key Laboratory of New Ceramics and Fine Processing, School of Materials Science and Engineering, Tsinghua University, Beijing 100084, China}
\author{Kai Liu}
\email[]{kliu@ruc.edu.cn}
\affiliation{Department of Physics and Beijing Key Laboratory of Opto-electronic Functional Materials $\&$ Micro-nano Devices, Renmin University of China, Beijing 100872, China}
\affiliation{Key Laboratory of Quantum State Construction and Manipulation (Ministry of Education), Renmin University of China, Beijing 100872, China}
\author{Xu-Cun Ma}
\email[]{xucunma@mail.tsinghua.edu.cn}
\affiliation{State Key Laboratory of Low-Dimensional Quantum Physics, Department of Physics, Tsinghua University, Beijing 100084, China}
\affiliation{Frontier Science Center for Quantum Information, Beijing 100084, China}
\author{Qi-Kun Xue}
\email[]{qkxue@mail.tsinghua.edu.cn}
\affiliation{State Key Laboratory of Low-Dimensional Quantum Physics, Department of Physics, Tsinghua University, Beijing 100084, China}
\affiliation{Department of Physics, Southern University of Science and Technology, Shenzhen 518055, China}
\affiliation{Frontier Science Center for Quantum Information, Beijing 100084, China}
\affiliation{Hefei National Laboratory, Heifei 230088, China}
\affiliation{Beijing Academy of Quantum Information Sciences, Beijing 100193, China}
\affiliation{Quantum Science Center of Guangdong-Hongkong-Macao Greater Bay Area (Guangdong), Shenzhen, China}
\author{Can-Li Song}
\email[]{clsong07@mail.tsinghua.edu.cn}
\affiliation{State Key Laboratory of Low-Dimensional Quantum Physics, Department of Physics, Tsinghua University, Beijing 100084, China}
\affiliation{Frontier Science Center for Quantum Information, Beijing 100084, China}

\begin{abstract}
Interfacial interactions often promote the emergence of unusual phenomena in two-dimensional systems, including high-temperature superconductivity. Here, we report the observation of full-gap superconductivity with a maximal spectroscopic temperature up to 26 K in a BaAs monolayer grown on ferropnictide Ba(Fe$_{1-x}$Co$_x$)$_2$As$_2$ (abbreviated as BFCA) epitaxial films. The superconducting gap remains robust even when the thickness of underlying BFCA is reduced to the monolayer limit, in contrast to the rapid suppression of $T_\textrm{c}$ in standalone BFCA thin films. We reveal that the exceptional crystallinity of the BaAs/BFCA heterostructures, featured by their remarkable electronic and geometric uniformities, is crucial for the emergent full-gap superconductivity with mean-field temperature dependence and pronounced bound states within magnetic vortices. Our findings open up new avenues to unravel the mysteries of unconventional superconductivity in ferropnictides and advance the development of FeAs-based heterostructures.

\end{abstract}

\maketitle
\begin {spacing}{0.989}
Despite decades of intense research, the unconventional pairing mechanism in high-temperature ($T_\textrm{c}$) cuprate and iron-based superconductors remains among the most challenging and unresolved issues in modern condensed matter physics. In these materials, superconductivity is commonly induced by suppressing a proximate antiferromagnetic order through chemical doping \cite{lee2006doping,johnston2010puzzle}, which inevitably introduces randomness and can even change the crystal structure. Alternatively, two-dimensional (2D) interfaces mimic the layered structure of high-$T_\textrm{c}$ superconductors and provide promising platforms for achieving high-$T_\textrm{c}$ superconductivity. Meanwhile, these interfaces enable a spatial balance between dissipationless macroscopic transport associated with Cooper pairs, and the disorder introduced by chemical doping that breaks electron pairing \cite{song2020emergent}. A well-known example of this strategy is FeSe monolayers grown on SrTiO$_3$ substrates, where superconductivity is enhanced by an order of magnitude \cite{wang2012interface, ge2015superconductivity}. Recent advancements in material engineering have allowed for the experimental realization of 2D superconductivity in various systems, such as oxide heterostructures \cite{gozar2008high, reyren2007superconducting, liu2021two, chen2021electric}, epitaxial crystalline films \cite{zhang2008superconductivity, xi2016ising, yi2024interface}, gated crystal surfaces \cite{ueno2008electric, ueno2011superconductivity}, and magic-angle graphene superlattices \cite{cao2018unconventional, chen2019signatures}.

Among these systems, oxide heterostructures involving cuprates and iron selenides exhibit the highest $T_\textrm{c}$ on record, making them highly intriguing for both fundamental research and potential applications \cite{wang2012interface, ge2015superconductivity,gozar2008high,logvenov2009high,shen2023reentrance}. However, designing and growing clean and superconducting 2D heterostructures with high crystallinity, tunability and $T_\textrm{c}$ remains an unprecedented challenge in condensed matter physics. In this study, we report the realization of such heterostructures by growing a crystalline BaAs monolayer on ferropnictide Ba(Fe$_{1-x}$Co$_x$)$_2$As$_2$ (BFCA) films. The heterostructure exhibits robust full-gap superconductivity with a maximum $T_\textrm{c}$ up to 26 K, as spectroscopically revealed by high-resolution scanning tunneling microscopy (STM).

Bulk BFCA represents the prototypical electron-doped ferropnictide superconductor with a maximum $T_\textrm{c}$ of 22 K \cite{sefat2008superconductivity}. Here we instead prepared epitaxial BFCA thin films on SrTiO$_3$(001) substrates by using molecular beam epitaxy (MBE), as described in the Supplemental Materials \cite{Supplementary}. High quality of as-grown BFCA films with a recorded onset $T_\textrm{c}$ of 29 K  \cite{li2017improving, canfield2010feas} and atomically sharp TiO$_2$-BFCA interface were demonstrated by X-ray diffraction (XRD), high-resolution transmission electron microscopy (TEM), and electrical transport [Supplementary Section 1]. To enable STM measurements, BFCA films were prepared on 0.5-wt$\%$ Nb-doped SrTiO$_3$(001) substrates as well. Owing to the negative charge of the superconducting (Fe,Co)As plane, as-grown BFCA invariably exhibits Ba-terminated surfaces with two distinct Ba coverages, namely a superconducting 1/4 Ba surface with a quasi-2 $\times$ 2 reconstruction and a metallic 1/8 Ba surface with a 2$\sqrt{2}$ $\times$ 2$\sqrt{2}R45^\circ$ reconstruction [Supplementary Section 1].
\end {spacing}

\begin{figure}[t]
\includegraphics[width=\columnwidth]{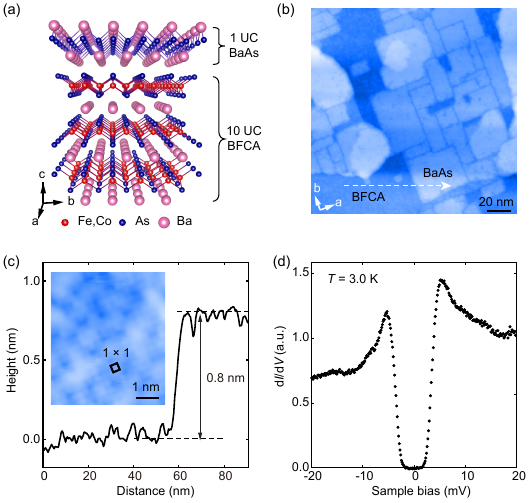}
\caption{(a) Schematic structure of the BaAs/BFCA heterostructure. (b) Typical STM topography (175 nm $\times$ 175 nm, $V = 3$ V, $I = 5$ pA) of BaAs films prepared on BFCA. The white arrows mark the crystalline direction of BFCA in the tetragonal phase. (c) Height profile measured along the dashed line in (b), revealing a height difference of $\sim$ 0.8 nm between BaAs and the underlying BFCA. Inset shows the atomically resolved image of monolayer BaAs. (d) Low-energy scale $dI/dV$ spectra showing a full-gap superconducting gap centered at $E_\textrm{F}$.
}
\end{figure}

\begin{figure}[h]
\includegraphics[width=\columnwidth]{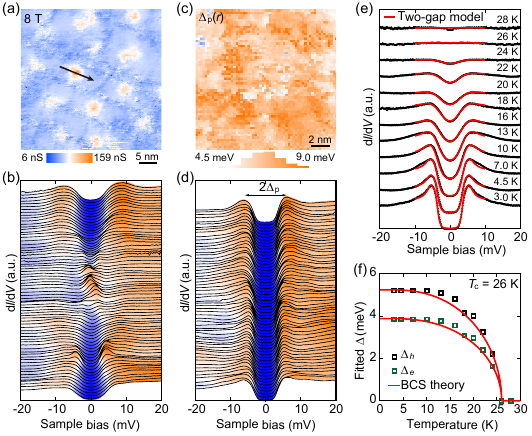}
\caption{(a) Typical ZBC map showing Abrikosov vortices in an 8 T external magnetic field. The ZBC map at zero field was subtracted as a background to eliminate the electronic inhomogeneity caused by BaAs cracks. (b) Spatially resolved $dI/dV$ spectra measured along the arrowed line in (a), exhibiting prominent CdGM states within the vortices. (c) Gap map $\Delta_\textrm{p}(r)$ acquired over a field of view of 15 nm $\times$ 15 nm with a resolution of 60 pixels $\times$ 60 pixels. (d) A series of linecut $dI/dV$ spectra measured along a 15 nm trajectory, illustrating the high homogeneity of the superconducting gap. (e) Evolution of the full-gap superconductivity with increasing temperatures from 3 K to 28 K. Red curves show the best fits of the experimental data to a two-gap Dynes model. (f) Fitted gap sizes $\Delta_{e,h}$ as a function of temperature, revealing a mean-field dependence (red curves). All experimental data were taken on the optimally doped BaAs/BFCA heterostructure.
}
\end{figure}

High-crystalline heterostructures of BaAs/BFCA [Fig.\ 1(a)] were then synthesized by depositing additional Ba onto the as-grown BFCA films, followed by sequential annealing under an As atmosphere to convert Ba into BaAs. As shown in Fig.\ 1(b), the resultant BaAs monolayers are featured by flat 2D rectangular islands along either of the As-As bonding directions of the underlying BFCA. Careful measurements reveal a step height of approximately 0.8 nm [Fig.\ 1(c)] and a square lattice with an in-plane constant of $\sim$ 3.9 \AA\ for BaAs (inset). The in-plane lattice mismatch ($<0.1$ \AA) between BaAs and BFCA is negligible. These lattice parameters diverge from all previously reported Ba$_m$As$_n$ structures but align with the theoretically predicted $\alpha$-PbO-type BaAs \cite{shim2009density}. This BaAs motif was previously proposed as a spacer layer for BaFeAs$_2$, isostructural to the 1111-type ferropnictides \cite{kamihara2008iron}. Indeed, the experimental observation of two Ba atoms per unit cell is consistent with the structure anticipated for the $\alpha$-PbO-type BaAs [Supplementary Section 2]. Moreover, our density functional theory (DFT) calculations indicate higher stability of BaAs compared to the 1/2Ba adlayer on as-grown BFCA, further supporting our identification [Supplementary Section 3].

Unexpectedly, the BaAs/BFCA heterostructures show full-gap superconductivity with pronounced coherence peaks and vanishing conductance within a finite energy range around the Fermi energy ($E_\textrm{F}$), as illustrated in Fig.\ 1(d). The superconducting nature of the U-shaped energy gap is confirmed by our observation of Abrikosov vortices under external magnetic fields, a key hallmark of type-II superconductors \cite{abrikosov1957magnetic}. Figure 2(a) displays a representative zero-bias conductance (ZBC) map measured under a magnetic field of 8 T perpendicular to the BaAs/BFCA heterostructure (see Supplementary Section 4 for more maps at various magnetic fields). As expected, the magnetic vortices appear as regions of enhanced ZBC extending over 2 $\sim$ 3 nm and increase in number with the applied magnetic field. More importantly, within individual vortices away from the BaAs cracks [Fig.\ 2(a)], a prominent ZBC peak emerges and reduces in strength with increasing distance from the vortex center [Fig.\ 2(b)]. This behavior is characteristic of Caroli-de Gennes-Matricon (CdGM) bound states within vortices of a clean $s$-wave superconductor, indicating minimal disorder effects \cite{hayashi1998low, renner1991scanning}. The weak scattering is consistent with the small spatial fluctuation of the gap magnitude $\Delta_\textrm{p}(r)$, measured as half of the energy distance between the two coherence peaks, and the absence of pair-breaking-induced low-lying excitations inside the superconducting gap [Figs.\ 2(c) and 2(d)]. By fitting the $\Delta_\textrm{p}(r)$ probability distribution with a Gaussian function [Fig.\ 2(c)], we determined $\Delta_\textrm{p}$ and its standard deviation $\sigma$ to be 6.8 meV and 1.2 meV, respectively.

To shed light on the observed full-gap superconductivity, we attempted to fit the gap using a phenomenological Dynes model \cite{dynes1978direct}. It was found that a single-gap scenario fails to simultaneously account for the flat bottom and broadened coherence peaks of the superconducting gap, particularly at low temperatures [Supplementary Section 5]. Considering the multi-Fermi sheet nature of BFCA \cite{terashima2009fermi, sekiba2009electronic}, we employed a two-gap Dynes formula \cite{teague2011measurement}
\begin{equation}
\frac{dI}{dV} = \sum_{i=e,h} B_i \int \text{Re}\left[\frac{(E - i\Gamma)}{\sqrt{(E - i\Gamma)^2 - \Delta^2}}\right] \left(\frac{\partial f}{\partial E}\right)dE,
\end{equation}
where $\Delta_e$ and $\Delta_h$ represent the superconducting gaps in the electron and hole pockets \cite{terashima2009fermi, sekiba2009electronic}, respectively. Detailed fitting procedures and parameters are described in Supplementary Section 5. By considering two isotropic gaps with a fixed gap ratio ($\Delta_h/\Delta_e = 1.35$) \cite{terashima2009fermi} and spectral weights ($B_e = 0.7$, $B_h = 0.3$) on the electron and hole pockets, respectively, we can reproduce the $dI/dV$ spectra at various temperatures after subtracting a sloped background, as shown in Fig.\ 2(e). This model involves no gap anisotropy or nodes, which aligns well with the full-gap $s$-wave superconductivity widely discussed in the context of ferropnictides \cite{mazin2008unconventional,kuroki2008unconventional,hoffman2011spectroscopic, hirschfeld2011gap, fernandes2022iron}. Figure 2(f) shows the extracted $\Delta_e$ and $\Delta_h$, both of which display a mean-field temperature dependence with a transition temperature, $T_\textrm{c}$, of approximately 26 K. The gap ratios $2\Delta/k_B T_c \sim 3.4$(4.6) for $\Delta_e$($\Delta_h$) conform to the BCS prediction ($\sim 3.53$) in the weak (strong)-coupling limit. Notably, our simulation yields a small Dynes parameter of $\Gamma_{e,h} \sim 0.08$ meV, nearly an order of magnitude smaller than those in cleaved BFCA surfaces \cite{teague2011measurement}, and comparable to the stoichiometric FeSe superconductor with a highly ordered surface \cite{liu2018robust}. This finding compellingly indicates minimal disorder scattering in the BaAs/BFCA heterostructures, driving the system into the clean limit.

Next, we investigated the evolution of the superconductivity in BaAs/BFCA (OP, $x \sim 0.061$) with the BFCA thickness. As revealed in Fig.\ 3(a), the superconducting gap magnitude ($\Delta_\textrm{p}$) remains nearly constant despite minor gap filling, in stark contrast to the rapid decline of the onset transition temperature ($T_\textrm{c}$) in as-grown BFCA films at the identical Co doping level [Figs.\ 3(b) and 3(c)]. For the standalone BFCA films, the $T_\textrm{c}$ gradually decreases to 18 K in 2 UC films, representing a reduction of around 40\% compared to that of 10 UC films. This reduction may be potentially attributed to the intrinsic suppression of the long-range superconducting order at the 2D limit \cite{mermin1966absence}, compressive strain, and/or interfacial disordering from the underlying STO substrates [Supplementary Section 1]. Furthermore, the full-gap superconductivity can be tuned by Co chemical doping in the underlying BFCA films. Figure 4(a) represents the $dI/dV$ spectra of several BaAs/BFCA (10 UC) heterostructures at different doping levels. It is apparent that the superconducting energy gaps reduce in magnitude as the Co doping level deviates from the optimal value at $x = 0.061$ and exhibit a domed superconducting phase diagram, as shown in Fig.\ 4(b).

\begin{figure}[h]
\includegraphics[width=1.0\columnwidth]{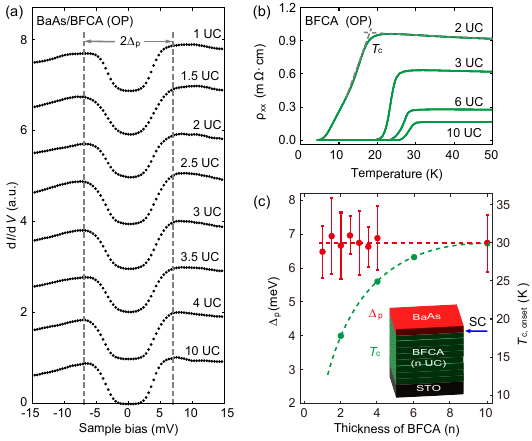}
\caption{(a) Spatially averaged superconducting energy gaps in BaAs/BFCA (OP, $x \sim 0.061$) heterostructures with varied BFCA thickness as labeled. (b) Thickness dependence of the in-plane resistivity $\rho_{ab}$ of as-grown BFCA films on undoped SrTiO$_3$. (c) Averaged gap magnitude $\Delta_\textrm{p}$ measured on BaAs/BFCA and transport $T_{\textrm{c}}$ of standalone BFCA as a function of the BFCA film thickness, highlighting the importance of BaAs/FeAs interface (sketched in the inset) in the observed full-gap superconductivity. The error bars in $\Delta_\textrm{p}$ correspond to standard deviations from multiple measurements.
}
\end{figure}

\begin{figure}[t]
\includegraphics[width=\columnwidth]{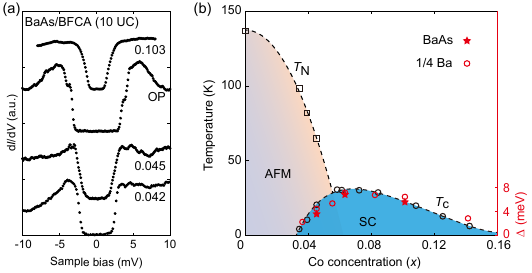}
\caption{(a) Representative $dI/dV$ spectra measured at 0.4 K on BaAs/BFCA heterostructures with varying Co doping levels. (b) Phase diagram of 10 UC BaAs-BFCA hetereostructures (red stars). The antiferromagnetic transition temperature $T_\textrm{N}$ (black square), $T_\textrm{c}$ (black circle) and the superconducting gap ($\Delta_\textrm{p}$) measured on the 1/4 Ba surface of as-grown BFCA are integrated for comparison. Here $T_\textrm{N}$ and $T_\textrm{c}$ were determined from the Co doping-dependent resistivity curves of BFCA films [Supplementary Section 1].}
\end{figure}

To clarify the mechanism underlying the observed full-gap superconductivity, we compare the surface geometry and the electronic states of BaAs/BFCA and the 1/4 Ba-terminated BFCA (abbreviated as 1/4 Ba-BFCA), which consistently display V-shaped  superconducting gaps with considerable in-gap excitations near $E_\textrm{F}$. We note that the 1/4 Ba surfaces exhibit many absorbed adatoms that must lead to significant scattering on the quasi-2 $\times$ 2 superstructure, in sharp contrast to the highly-ordered BaAs monolayer [Fig. 1(c) and Supplementary Section 1]. The enhanced crystallinity of BaAs/BFCA is further demonstrated by the reduced energy fluctuation of the characteristic density of state (DOS) peaks in the wide-energy-scale $dI/dV$ spectra and the residual conductance at $E_\textrm{F}$ [Supplementary Section 5]. Thus, we conclude that the full-gap superconductivity is benefited from the high crystallinity of the BaAs/BFCA heterostructures, which minimizes pair-breaking effects and results in a small Dynes parameter $\Gamma$. In contrast, the large spatial inhomogeneity can lead to a significant scattering rate $\Gamma$ or even an energy-dependent $\Gamma(E)$ \cite{zou2017effect,shan2011observation,alldredge2008evolution}. Indeed, by incorporating a large $\Gamma_h = 1.2$ meV or $\Gamma_e = \alpha E$ ($\alpha \sim 0.6 - 2.4$), the V-shaped gaps on the Ba-BFCA surface and their temperature evolution can be nicely reproduced [Supplemental Section 5]. Throughout these simulations, the gap magnitude $\Delta_{e,h}$, spectral weights $B_{e,h}$, and $T_\textrm{c}$ are essentially unchanged, except for the variation in the disorder scattering rate $\Gamma$.

Our results provide crucial insights into the unconventional superconductivity in ferropnictides, with implications for high-$T_\textrm{c}$ superconductors. Firstly, the superconductivity most likely originates from FeAs-derived 2D interfaces and remains full-gap as long as both sides of the interface are sufficiently clean to minimize pair-breaking effects. In the BaAs/BFCA heterostructures constructed here, the high crystallinity of BaAs ensures a potential dissipationless channel for the full-gap superconductivity in the clean limit, despite some scattering induced by in-plane Co dopants \cite{zou2017effect,zhou2011electronic,yin2009scanning,nishizaki2011surface}. However, pair-breaking is inevitable in the 1/4 Ba-BFCA systems due to the occurrence of disorder within both the Ba surface and the (Fe, Co)As plane, leading to a V-shaped superconducting gap with significant in-gap states. This phenomenological model explains the mysterious V-shaped gaps previously observed in cleaved BFCA surfaces with substantial disorder \cite{terashima2009fermi,hoffman2011spectroscopic,alldredge2008evolution,zhou2011electronic}. Second, our observation of the CdGM states within magnetic vortices highlights weak impurity scattering from the in-plane Co doping. We therefore speculate that the absence of vortex bound states in earlier STM measurements of cleaved bulk crystals \cite{yin2009scanning,nishizaki2011surface} was also derived from strong scattering from disordered surfaces. The weak scattering by in-plane Co is consistent with the broad superconducting dome spanning a wide range of Co doping levels, in stark contrast to the rapid $T_\textrm{c}$ suppression seen with Cu substitution in cuprates \cite{tarascon1987metal}.

Lastly, we observe that the superconducting gap magnitude $\Delta_\textrm{p}$ and its evolution with Co doping level $x$ remain consistent for both BaAs/BFCA heterostructures and 1/4 Ba-BFCA surfaces [Fig.\ 4(b)]. This not only underscores the important role of disorder in shaping the superconducting gap structure, but also suggests similar effective electron doping in both systems. To verify this, we carried out DFT calculations on the carrier density of the topmost FeAs plane with different coating layers (Supplementary Section 3). Using the 1/2 Ba-terminated surface as a reference ($n_\textrm{2D} \sim$ 0), the relative carrier density was calculated to be $n_\textrm{2D} \sim$ 0.01 electrons/Fe for the BaAs layer and 0.02  $\sim$ 0.04 holes/Fe for the 1/4Ba layer, respectively. The termination-induced doping variations, particularly between the 1/2Ba and BaAs-terminated layers, are relatively modest compared to those from in-plane Co doping \cite{sekiba2009electronic}, explaining the anchoring superconductivity behavior observed on 1/4 Ba and BaAs surfaces with the transport $T_\textrm{c}$ [Fig. 4(b)]. Furthermore, we noted that the Fe 3$d_{z^2}$ orbital-derived DOS peaks on the BaAs surface (Supplementary Section 3 and Supplementary Section 5) aligns its energy well with that of bulk BFCA \cite{zhang2011orbital}. This makes the BaAs/BFCA heterostructures a promising clean system to investigate the intrinsic properties of ferropnictide superconductors. At the same time, the DOS contributions from Fe $d_{xz}/d_{yz}$ orbitals surpass those of the Fe $d_{z^2}$ orbital near $E_\textrm{F}$ on the BaAs surface, indicating significant alterations in the Fe 3$d$ orbital distributions with different surface layers. These variations impact the topology of the Fermi surface, a crucial ingredient for unusual electron pairing in multi-orbital superconductors.

In summary, we have designed and constructed highly crystalline BaAs/BFCA heterostructures with high $T_\textrm{c}$ and tunable carrier density. The heterostructures exhibit full-gap superconductivity with significant spatial homogeneity and prominent CdGM bound states within vortices. Our results reveal that the BaAs adlayer serves as an ideal template to reveal the intrinsic superconductivity of ferropnictides by suppressing disorder scattering while maintaining the carrier density. Considering the structural similarity between BaAs/BFCA interface and the 112 family of iron-based superconductors \cite{Katayama2013CaLaFeAs2}, this interface may provide a novel system for exploring topological superconductivity \cite{Wu2015CaFeAs2,Wu2015AsChain}. In addition, the high-quality BFCA epitaxial films, with significant tunability in thickness, chemical doping and surface terminations, provide promising platforms for developing FeAs-based heterostructures and devices.

\begin{acknowledgments}
This work was financially supported by the National Natural Science Foundation of China (12141403, 12174443, 12304164, 12134008), the National Key R\&D Program of China (2022YFA1403100, 2024YFA1408100), the Innovation Program for Quantum Science and Technology (2021ZD0302502) and the Shenzhen Natural Science Foundation (JCYJ20240813094203005).

M.\ Q.\ Ren and Q.\ J.\ Cheng contributed equally to this work.

\end{acknowledgments}

%

\end{document}